\documentclass{nature}
\usepackage{booktabs}
\usepackage{graphicx}
\usepackage{dcolumn}
\usepackage{bm}
\usepackage{amsmath,amssymb}
\usepackage{siunitx}
\usepackage{caption}

\usepackage{xcolor}

\def\nar{\ref@jnl{New A Rev.}}          
                
\def\msun{$M_\odot$}

\usepackage{graphicx}
\makeatletter
\let\saved@includegraphics\includegraphics
\AtBeginDocument{\let\includegraphics\saved@includegraphics}
\renewenvironment*{figure}{\@float{figure}}{\end@float}
\makeatother
\begin{document}

\title{A 2 per cent Hubble constant measurement from standard sirens within 5 years
}

\author{Hsin-Yu Chen$^{1,2,*}$, Maya Fishbach$^2$ and Daniel E. Holz$^{2,3,4}$}

\maketitle

\begin{affiliations}
 \item Black Hole Initiative, Harvard University, Cambridge, Massachusetts 02138, USA
 \item Department of Astronomy and Astrophysics, University of Chicago, Chicago, Illinois 60637, USA
 \item Enrico Fermi Institute, Department of Physics, and Kavli Institute for Cosmological Physics, University of Chicago, Chicago, Illinois 60637, USA
 \item Physics Department and Kavli Institute for Particle Astrophysics \& Cosmology, Stanford University, Stanford, CA 94305
\end{affiliations}

\begin{abstract}
Gravitational wave coalescence events provide an entirely new way to determine the Hubble constant~\cite{Schutz1986,Holz2005,2006PhRvD..74f3006D}, with the absolute distance calibration provided by the theory of general relativity. This standard siren method was utilized to measure the Hubble constant using LIGO-Virgo's detection of the binary neutron-star merger GW170817, as well as optical identifications of the host galaxy, NGC 4993~\cite{2017Natur.551...85A}. 
The novel and independent measurement is of particular interest given the existing tension between the value of the Hubble constant determined using Type Ia supernovae via the local distance ladder ($73.24 \pm 1.74$) and that from Cosmic Microwave Background observations ($66.93 \pm 0.62$) by $\sim 3$ sigma \cite{SHoES,PlanckCosmology}. Local distance ladder observations may achieve a precision of 1\% within 5 years, but at present there are no indications that further observations will substantially reduce the existing discrepancies \cite{Riess:2018}.
In addition to clarifying the discrepancy between existing low and high-redshift measurements, a precision measurement of the Hubble constant is of crucial value in elucidating the nature of the dark energy~\cite{2005ASPC..339..215H,2018arXiv180607463D}.
Here we show that LIGO and Virgo can be expected to constrain the Hubble constant to a precision of $\sim2\%$ within 5 years and $\sim1\%$ within a decade.
\end{abstract}

We explore the expected constraints on the Hubble constant ($H_0$) from gravitational-wave standard sirens. The gravitational-wave data provides a direct measurement of the luminosity distance to the source, but the redshift must be determined independently. We consider gravitational-wave events both with (``counterpart'') and without (``statistical'') direct electromagnetic measurements of the source redshift, and carry out an end-to-end simulation of the $H_0$ measurement from a simulated data set consisting of 30,000 binary neutron star (BNS) mergers and 60,000 binary black hole (BBH) mergers. We include realistic measurement uncertainties, galaxy peculiar velocities, and selection effects in our analysis. 

We anticipate that most, if not all, binary neutron star mergers detected in gravitational waves will have an electromagnetic counterpart 
(e.g. from associated isotropic~\cite{1998ApJ...507L..59L,2010MNRAS.406.2650M} kilonova emission~\cite{2017Sci...358.1556C,2017ApJ...848L..16S}) 
that will allow for a unique host galaxy identification~\cite{2017Natur.551...85A}. Assuming the BNS population is similar to the population of short gamma ray bursts, 
we expect the typical offset between a kilonova and its associated host galaxy to be no more than 100 kpc~\cite{2013ApJ...776...18F}. Since Advanced LIGO-Virgo BNS detections 
will be within 400 Mpc, it will be possible to identify host galaxies down to 0.003 $L^\star_B$ (apparent magnitudes $< 23$) with modest observational resources. We find 
that in this counterpart case, the fractional $H_0$ uncertainty will scale roughly as  $15\%/\sqrt{N}$, where $N$ is the number of BNS mergers detected by the LIGO-Hanford, LIGO-Livingston, and Virgo network (HLV).
Throughout, we quote fractional $H_0$ measurement uncertainties defined as half the width of the symmetric 68\% credible interval divided by the median. 
If KAGRA and LIGO-India join the detector network (HLVJI), this convergence improves slightly to $13\%/\sqrt{N}$, as a five-detector network tends to provide better measurements of the source inclination, and therefore distance, due to the improved polarization information.  

We note that the representative $\sigma_{h_0}$ (15\% for the three-detector network, and 13\% for the five-detector network), is smaller than the typical width of the $H_0$ measurement from an individual event (GW170817 provided an unusually tight measurement; see Extended Data). This is due to the fact that for a single event, the $H_0$ posterior probability density function is a highly non-Gaussian function; the distance-inclination degeneracy leads to long tails to large distances (and low $H_0$ values) for edge-on sources, and tails in the opposite direction for face-on sources. Combining these asymmetric distributions leads to a $1/\sqrt N$ convergence with a smaller effective $\sigma_{h_0}$ than the width of a typical single-event $H_0$ measurement ~\cite{2010PhRvD..81l4046H,2013arXiv1307.2638N}. Furthermore, due to the asymmetry of the single-event measurements, it may take $\sim20$ events to reach the expected $1/\sqrt{N}$ convergence rate.
For example, we may get lucky in the first few events and get an unusually good $H_0$ measurement (GW170817 is an excellent example of this), after which we will converge more slowly than $1/\sqrt{N}$ for some time as we detect average events. After $\sim20$ events, however, we have a sufficient statistical sample of detections to have converged to a representative $\sigma_{h_0}$ for the population. At this point, the combined $H_0$ measurement approaches a Gaussian distribution and we reach the expected $1/\sqrt{N}$ behavior. 

\begin{figure}
\includegraphics[width=\columnwidth]{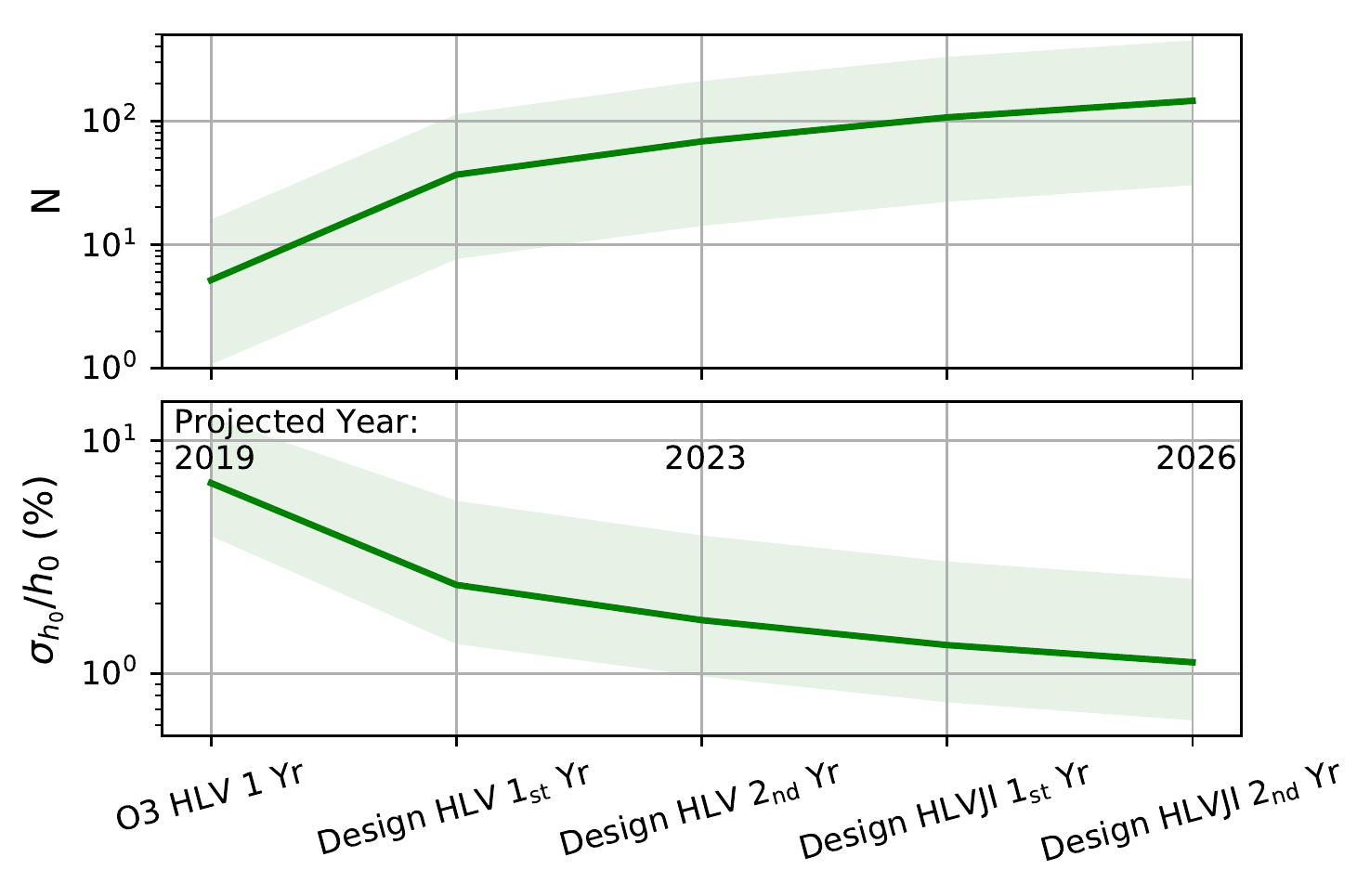}
\caption{\label{fig:twopanel}
 \textbf{Projected number of binary neutron-star detections and corresponding fractional error for the standard siren $H_0$ measurement} \emph{Top panel:} The expected total number of BNS detections for future observing runs, using the median merger rate (solid green curve), and upper/ lower rate bounds (shaded band). \emph{Bottom panel:} The corresponding $H_0$ measurement error, defined as half of the width of the 68\% symmetric credible interval divided by the posterior median. The band corresponds to the uncertainty in the merger rate shown in the top panel. These measurements assume an optical counterpart, and associated redshift, for all BNS systems detected in gravitational-waves. }

\end{figure}


\begin{figure}
\centering
\includegraphics[width=1.1\columnwidth]{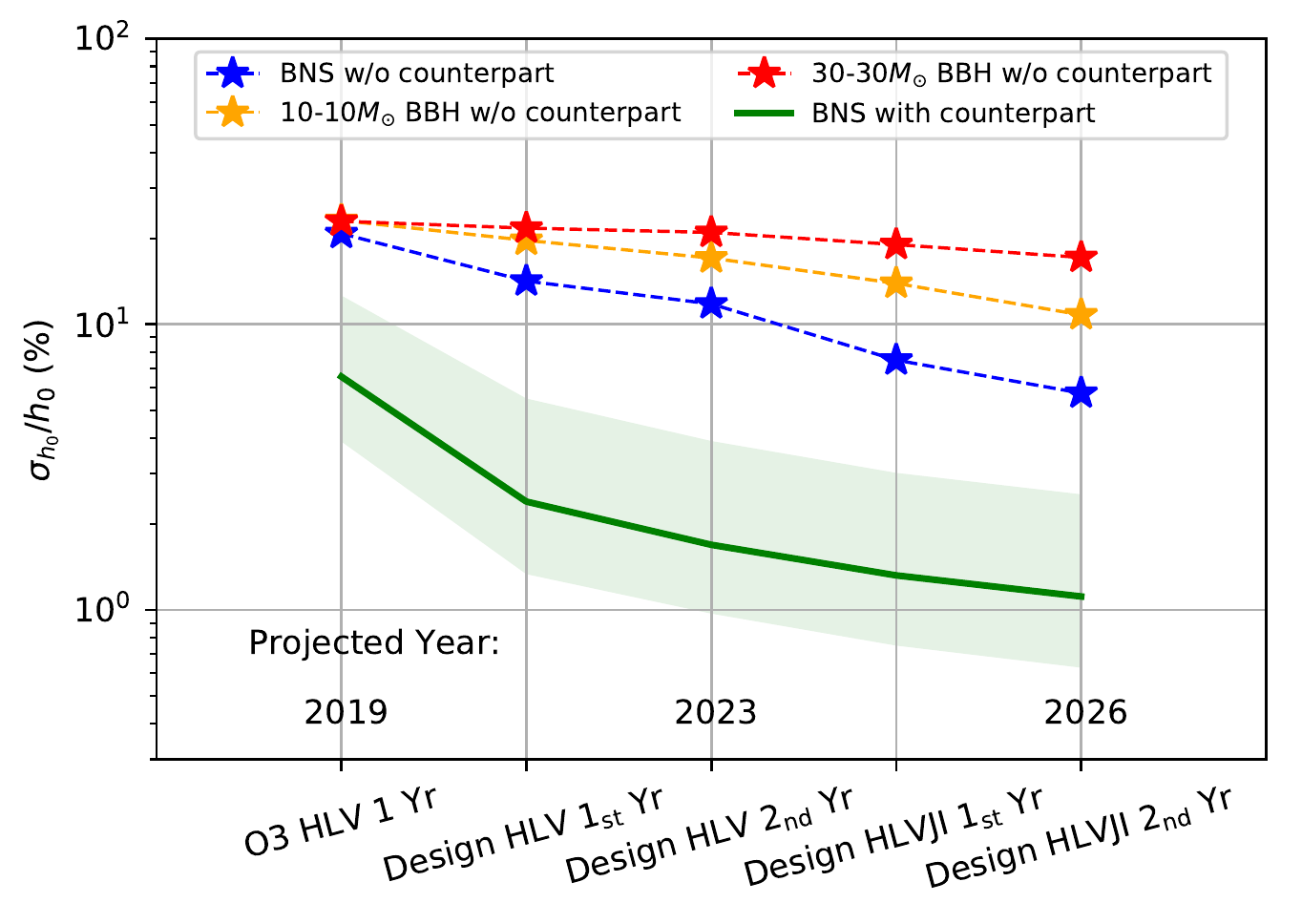}
\caption{\label{fig:obsstage} \textbf{ Projected fractional error for the standard siren $H_0$ measurement for binary neutron stars and binary black holes for future gravitational-wave detector networks.} The green shaded band (identical to the band in Figure~\ref{fig:twopanel}) corresponds to the BNS rate uncertainty; the same rate uncertainty applies to the `BNS without counterpart' curve.
For the `without counterpart' curves, we adopt a statistical standard siren approach using only events localized to within 10,000 Mpc$^3$ (90\% credible region); events with larger volumes do not contribute noticeably (see the text for more details).}
\end{figure}

In order to predict how the $H_0$ measurement improves with time, we consider the BNS rates inferred from GW170817\cite{Abbott:170817}, $1540^{+3200}_{-1220}$ Gpc$^{-3}$yr${^{-1}}$, together with the planned network sensitivity and duty cycle, to compute the expected number of detections at each observing stage. Figure~\ref{fig:twopanel} shows the improvement in the $H_0$ measurement as BNS detections with unique host galaxies are accumulated.
We start with a 15\% prior measurement on $H_0$, representing the constraint from GW170817~\cite{2017Natur.551...85A}; we approximate this by a Gaussian centered at 67.8 km/s/Mpc with a standard deviation of 10.2 km/s/Mpc, but the exact center and shape of the $H_0$ posterior do not affect our results.
The rate of detections will increase as the gravitational-wave network improves in sensitivity between LIGO-Virgo’s third observing run (`O3 HLV'), HLV at design sensitivity (`Design HLV'), and the five-detector network (`Design HLVJI'), from an average of 5 BNS detections per year in O3, to 32 and 39 detections per year for Design HLV and Design HLVJI. 
The merger rate provides the major source of uncertainty in predicting the $H_0$ measurement error. The solid line in the bottom panel of Figure~\ref{fig:twopanel} shows the average $H_0$ measurement error over 100 realizations assuming the median BNS merger rate, while the lower/ upper bounds of the shaded band assume the upper/ lower 90\% bounds on the merger rate inferred from GW170817.

We find that, if it is possible to independently measure a unique redshift for all BNS events, the fractional uncertainty on $H_0$ will reach 2\% (at the $1\sigma$ level) by the end of 2 years of HLV at design sensitivity ($\sim2023$; corresponding to $\mathcal{O}(50)$ events), sufficient to arbitrate the current tension between local and high-$z$ measurements of $H_0$.
After $\mathcal{O}(100)$ BNS events, GW standard sirens would provide a $1\%$ determination of $H_0$. This is expected to happen after $\sim2$ years of operation of the full HLVJI network ($\sim2026$), but given the rate uncertainties could happen many years later, or could happen as early as 2023.
 
Not all sources will have associated transient electromagnetic counterparts: we may fail to identify the counterparts to some binary neutron star mergers, and counterparts are not expected for binary black hole mergers. In the case where a unique counterpart cannot be identified, it is possible to carry out a measurement of the Hubble constant using the statistical approach. In this case, the redshifts of all potential host galaxies within the gravitational-wave three-dimensional localization region are incorporated, yielding an $H_0$ measurement that is inferior to the counterpart case, but still informative once many detections are combined. 
This means that, in the absence of a counterpart, only those gravitational-wave events with small enough localization volumes yield informative $H_0$ measurements. If the localization volume is too large, it contains a large number of potential host galaxies, significantly washing out the contribution from the correct host galaxy. Additionally, it may be difficult in practice to construct a complete galaxy catalog over a large volume with precise galaxy redshifts. 
We find that for BNSs without counterparts, combining the $H_0$ measurement from events that are localized to within 10,000 Mpc$^3$ ($\sim$40\% of events) yields identical constraints to the combined measurement utilizing the full sample -- events localized to greater than 10,000 Mpc$^3$ do not contribute to the measurement. For this reason, we use only the sources localized to within 10,000 Mpc$^3$ for the no-counterpart projections in Figure~\ref{fig:obsstage}. Note that for all of the no-counterpart curves in Figure~\ref{fig:obsstage}, we start with a flat $H_0$ prior between 50--100 km/s/Mpc. 

Because BBH systems tend to have much larger localization volumes than BNS systems (as they are more massive and found at greater distances), the statistical $H_0$ measurement for BBHs converges very slowly, even though they are detected at higher rates. We consider both ``light'' (10--10\,\msun) and ``heavy'' (30--30\,\msun) BBHs, assuming merger rates of $80^{+90}_{-70}$ Gpc$^{-3}$yr${^{-1}}$ for 10--10\,\msun BBHs  and $11^{+13}_{-10}$ Gpc$^{-3}$yr${^{-1}}$ for 30--30\,\msun BBHs \cite{Abbott:170104} (see the Methods section for details). Only $\sim$3\% of the light BBHs and $\sim$0.5\% of the heavy BBHs are localized to within 10,000 Mpc$^3$, meaning we only expect to detect $16^{+19}_{-14}$ well-localized BBHs by 2026.
This leads to a $\sim$10\% $H_0$ measurement with BBHs by 2026. 
We note that the constraints from statistical BBH standard sirens improve if the BBH rates are on the high end, as well as if the BBH mass function favors low masses.

For the projections in Figure~\ref{fig:obsstage}, we assumed that galaxies are distributed uniformly in comoving volume and that complete catalogs are available. 
If we incorporate the clustering of galaxies due to large-scale structure, the convergence rate in the statistical case improves by a factor of $\sim$2.5 (see Methods). 
Incorporating this large-scale structure effect, we find that it will still take more than $\sim 50$ binary neutron stars without a counterpart to reach a 6\% $H_0$ measurement, 
compared to only $<10$ binary neutron stars with a counterpart. Meanwhile, accounting for galaxy catalog incompleteness provides an additional source of uncertainty 
(see Eq.~\ref{eq:incompleteness} in the Methods), which can cancel out some of the improvement due to large-scale structure. For example, for a galaxy catalog completeness 
of $50\%$, the $H_0$ measurement would be degraded by a factor $\lesssim 2$ \cite{MDC}. Therefore, incorporating the effects of large-scale structure and catalog incompleteness, 
we expect that in practice the $H_0$ constraints in the statistical case will be {slightly better than our prediction in Figure~\ref{fig:obsstage}}, where the precise factor 
depends on properties of the relevant host galaxies and completeness of the catalog.  

While we have considered the counterpart and no-counterpart cases, we can also anticipate a situation in which we have a counterpart detection but no unambiguous host association. For example, an optical counterpart could be relatively isolated on the sky without a clearly identified host galaxy, or may have multiple possible host galaxies. In this case we can pursue a pencil-beam strategy focusing on the volume within $\sim100\,$kpc of the counterpart (see Methods). For BNSs, this will reduce the relevant volume to $\mathcal{O}(10)$ Mpc$^3$, for which we expect to have only $\sim 1$ potential host galaxy or galaxy group, thereby reducing to the counterpart case.

In addition to BNSs and BBHs discussed here, neutron star-black hole mergers are to be expected~\cite{Dominik:2015,Belczynski:2016b,2016Natur.534..512B}, and are likely to have detectable electromagnetic counterparts. Although the rates for these systems are uncertain and expected to be low, they will also be seen to greater distances than BNS systems, which may render them useful as standard sirens~\cite{2018arXiv180407337V}.

It is to be noted that our measurements of distance do not use any astrophysical modeling.
Alternatively, associated electromagnetic observations (for example from short gamma-ray burst afterglows or jet breaks) can provide additional constraints on the inclination, and thereby improve the individual measurements of $H_0$~\cite{2017Natur.551...85A,2017arXiv171006426G}. In this sense our counterpart results can be considered a conservative estimate. However, one of the advantages of standard sirens is that they are ``pure'' measurements of luminosity distance, avoiding complicated  astrophysical distance ladders or poorly understood calibration processes, and instead are calibrated directly by the theory of general relativity to cosmological distances. By introducing additional constraints based on astronomical observations (e.g., independent beaming measurements or estimates of the mass distribution or equation-of-state of neutron stars), there is the potential to introduce systematic biases that could fundamentally contaminate the standard siren measurements. In the present analysis we do not consider these additional constraints, although they may indeed have an important role to play in future standard siren science.

Eventually systematic errors in the amplitude calibration of the detectors may become a source of concern, as the luminosity distance is encoded in the amplitude of the gravitational-wave signal. However, the calibration uncertainty is currently limited by the photon calibrator to $\sim 1\%$, and this is likely to improve \cite{calibration}; we look forward to an era where sub-percent calibration becomes a necessity, but this is a number of years away. Another possible source of 
distance uncertainty is gravitational lensing. However, at the typical redshift of BNS and BBH ($z<0.5$ at design sensitivity) the effect will be minor relative to the uncertainty from the distance-inclination degeneracy~\cite{2005ApJ...631..678H}. In addition, for sufficient numbers of sources the effects of lensing will average away~\cite{2010PhRvD..81l4046H}. Of course, gravitational-wave cosmology is a new field, and unforeseen systematics could certainly arise as we push our measurements to the percent level and beyond.

We stress that our projected $H_0$ constraints are subject to several important uncertainties, the largest one of which is the  merger rate of BNS and BBH systems. The detection rates for BBHs depends sensitively on the mass distribution, which is not currently well-constrained \cite{2016ApJ...833L...1A,2017arXiv170908584F}.
Future detections will bring a better understanding of the merger rates and mass distributions of compact objects, allowing for improved predictions. Regardless, it is clear that gravitational-wave standard sirens will provide percent-level constraints on $H_0$ in the upcoming advanced detector era of gravitational-wave astronomy.

We acknowledge valuable and extensive discussions with Lindy Blackburn, Reed Essick, Will Farr, and Jonathan Gair. The authors were partially supported by NSF CAREER grant PHY-1151836 and NSF grant PHY-1708081. They were also supported by the Kavli Institute for Cosmological Physics at the University of Chicago through NSF grant PHY-1125897 and an endowment from the Kavli Foundation.
The authors acknowledge the University of Chicago Research Computing Center for support of this work.
HYC was supported in part by the Black Hole Initiative at Harvard University, through a grant from the John Templeton Foundation. MF was supported by the NSF Graduate Research Fellowship Program under grant DGE-1746045.

\section{Methods}
\label{sec:methods}
In what follows we present our method for inferring cosmological parameters from gravitational-wave (GW) and electromagnetic-wave (EM) measurements. We first Monte Carlo a representative sample of GW detections for a range of detector configurations. We then simulate the analysis of these data sets, and explore the resulting standard siren constraints. We highlight important aspects of our calculation, such as the role of peculiar velocities and selection effects.

\textbf{Synthetic Events and Host Galaxies} Measuring $H_0$ with standard sirens relies on our ability to extract the luminosity distance and sky position of GW sources. 
We follow the procedure in \cite{Chen:theone} to localize synthetic binary neutron star merger (BNS) and binary black hole merger (BBH) 
detections. The population of binaries are distributed uniformly in comoving volume in a Planck~\cite{PlanckCosmology} 
cosmology $(\Omega_{M_0}=0.308,\Omega_{\Lambda_0}=0.692,h_0=0.678)$. 
We assume the BNS merger rate follows the rate measured in~\cite{Abbott:170817}. To estimate the merger rate of 10--10\,\msun and 30--30\,\msun BBHs from the rate measured in~\cite{Abbott:170104}, we assume the BBH mass function follows a Salpeter power law and use 10--10\,\msun\ BBHs to characterize all BBHs with primary component masses between 5 and 15\,\msun, and 30--30\,\msun\ BBHs to characterize all BBHs with primary component masses between 20 and 50\,\msun. We do not place additional cuts on the secondary masses, which are distributed uniformly between 5\,\msun\ and the primary mass.

The \emph{detection}-rate of sources depends on the sensitivity, observing time, and duty-cycle of the GW detector network. We assume that the LIGO-Virgo network operates for one year at projected O3 sensitivity, followed by two one-year-long observing runs of LIGO-Hanford+ LIGO-Livingston+Virgo (HLV) at design sensitivity and two one-year-long runs of the five-detector network, LIGO-Hanford+ LIGO-Livingston+Virgo+KAGRA+LIGO-India (HLVJI), at design sensitivity \cite{Abbott:obs}. We take the combined duty cycle to be 0.5 for the HLV detector configuration and 0.3 for HLVJI. The number of detections is subject to Poisson statistics, and we simulate detections according to the merger rate, network sensitivity, observing time, and duty cycle.

To determine whether a binary merger is detected, we calculate the matched-filter signal-to-noise ratio (SNR) for each simulated binary. We draw the ``measured'' SNR from a Gaussian distribution centered at the matched-filter value with a standard deviation $\sigma=1$. Binary mergers are detected only if their measured network SNR is greater than 12. For each detected merger, we calculate its 3D localization according to the methods in \cite{Chen:theone} (We have verified that this procedure yields results which are consistent with the full parameter estimation pipeline, LALInference \cite{Veitch:2015}.) The 3D localization takes the form of a posterior probability distribution function, $p(\alpha,\delta,D_L | d_{\rm GW})$, over the sky position $(\alpha, \delta)$ and luminosity distance, $D_L$, given the GW data , $d_{\rm GW}$. 

The gravitational wave signal from each detected binary merger provides a measurement of $D_L$. To calculate $H_0$, we must also measure a redshift for each binary merger. Throughout, we take the redshift, $z$, to be the peculiar-velocity corrected redshift; that is, the redshift that the source would have if it were in the Hubble flow. We consider two cases: 
the redshift information either comes from a direct EM counterpart, such as a short gamma-ray burst/afterglow and/or a kilonova (``counterpart''), or a statistical analysis over a catalog of potential host galaxies (``statistical''). 

In the counterpart case, we assume that the EM counterpart is close enough to its host galaxy so that the host can be unambiguously identified, and we can measure its sky position and redshift. This is a reasonable assumption based on the distribution of offsets between short gamma ray bursts and their host galaxies, assuming short gamma ray bursts trace a similar population as BNS mergers, and taking into account that detected BNS mergers will be at much lower redshifts than the short GRB population. We assume that the sky position of each host galaxy is perfectly measured (i.e. with negligible measurement error), meaning we can fix the source sky position to the location of the counterpart in the GW parameter estimation (rather than marginalizing over all sky positions). The GW distance posterior changes slowly over the sky and therefore is not sensitive to the precise location of the counterpart. However, since the GW sky localization areas can be very large, fixing the source position can lead to important improvements in the distance, and hence $H_0$, measurements. We also assume that the peculiar-velocity corrected redshift, $z$, is measured with a 1-$\sigma$ error of (200 km/s)/$c$, which is a typical uncertainty for the peculiar velocity correction~\cite{Carrick:2015,Scolnic:2017b}. 

In the absence of an EM counterpart, we cannot identify a single host galaxy, and must use a catalog of all potential host galaxies~\cite{Schutz1986,2012PhRvD..86d3011D}. To simulate the galaxy catalogs, we consider two cases: a uniform-in-comoving-volume distribution of galaxies, and a distribution that follows the large-scale structure as simulated by the MICE galaxy catalog \cite{2015MNRAS.448.2987F,2015MNRAS.453.1513C,2015MNRAS.447.1319F,Carretero:2017zkw}. In the uniform distribution case, we construct a mock catalog by distributing galaxies uniformly in comoving volume with a number density of 0.02 Mpc$^{-3}$. This corresponds to the number density of galaxies 25\% as bright as the Milky Way, assuming the galaxy luminosity function is described by the Schechter function~\cite{1976ApJ...203..297S} with B-band parameters $\phi_{*}=1.6\times 10^{-2}h^3$\,Mpc$^{-3}$, $\alpha$=-1.07, $L_*=1.2\times 10^{10}h^{-2}L_{B,\odot}$,
and $h=0.7$ (where $L_{B,\odot}$ is the solar luminosity in B-band), and integrating down to $0.16\,L_*$ to find the luminosity density. (This corresponds to 83\% of the total
luminosity~\cite{2002MNRAS.336..907N,2003MNRAS.344..307L,2006A&A...445...51G}.) The lower luminosity limit of the MICE catalog is similar.
Thus, we assume that only galaxies brighter than $0.16 L_*$ can host binary mergers, although we note that the population of host galaxies is currently uncertain, and we can modify the assumed luminosity limit by including the effects of catalog incompleteness. A lower luminosity limit would increase the galaxy density and weaken the $H_0$ constraints in the statistical case.

\noindent\textbf{$H_0$ uncertainty} Not all GW events contribute equally to the $H_0$ measurements. In the counterpart case, the fractional error on the $H_0$ measurement from a single source depends on the fractional distance uncertainty of the GW source and the fractional 
redshift uncertainty of its host galaxy. To first order, this is:
\begin{equation}
\label{eq:counterr}
\left.\left( \frac{\sigma_{H_0}}{H_0} \right)^2 \right\vert_\mathrm{1 gal} \approx \left( \frac{\sigma_{v_H}}{v_H} \right)^2 + \left( \frac{\sigma_{D_L}}{D_L} \right)^2,
\end{equation}
where $v_H$ is the peculiar-velocity corrected ``Hubble velocity''. Because the recessional velocity uncertainty, 
$\sigma_{v_H}$, is typically around 150--250 km/s, the fractional recessional velocity decreases with distance. Meanwhile, the fractional distance uncertainty scales roughly inversely with SNR, and therefore tends to increase with distance. There is thus a ``sweet spot'', where the peculiar velocities and the distance uncertainties are comparable; for LIGO-Virgo's second observing run, this was $\sim30\,$Mpc, near the distance of GW170817. The sweet spot will increase as the networks become more sensitive; for detectors at design sensitivity the ideal BNS standard siren distance will be $\sim50\,$Mpc. At distances beyond this, the distance uncertainty will tend to dominate the peculiar velocity uncertainty; in this regime, the nearest (highest SNR) events tend to provide the tightest $H_0$ constraints. This can be seen in the figure in Extended Data, which shows the fractional $H_0$ uncertainty for individual events, plotted against the median posterior distance and 90\% posterior localization volumes. However, we note that the relationship between median distance, localization volumes, and fractional $H_0$ uncertainty is not very tight. Prior to identifying the counterpart for a particular event, we can estimate the accuracy of the $H_0$ measurement from the width and central value (e.g. median) of the GW distance posterior according to Eq.~\ref{eq:counterr}, using an estimated $v_H \approx 70\,\langle D_L \rangle$ km/s/Mpc, where $\langle D_L \rangle$ is the median GW distance. (Here we must use the GW posterior marginalized over the sky position, as we do not yet know the sky position of the counterpart.) We verify that this estimate of the combined distance and redshift uncertainty is a reasonable proxy for the resulting $H_0$ uncertainty, assuming an EM counterpart is found and provides an independent measurement of redshift. 

In the absence of a counterpart, we cannot assign a unique host, and so the $H_0$ error increases with the number of potential host galaxies in the localization volume. Significant galaxy clustering can mitigate this, as we discuss in the main text. For example, in the case of GW170817, the optical counterpart was found in NGC 4993, which is a member of a group of $\sim20$ galaxies, all of which have an equivalent Hubble recessional velocity \cite{2017ApJ...843...16K}. On the other hand, catalog incompleteness degrades the $H_0$ measurement, as we have to consider an additional background of uniformly distributed galaxies (see Equation~\ref{eq:incompleteness}).

\noindent\textbf{Bayesian Model} For a single event with GW and EM data, $d_{\rm GW}$ and $d_{\rm EM}$, we can write the likelihood as:
\begin{equation}
\label{eq:likelihood}
p(d_{\rm GW}, d_{\rm EM} | H_0) = \frac{\int p(d_{\rm GW}, d_{\rm EM}, D_L, \alpha, \delta, z | H_0) dD_L d\alpha d\delta dz}{\beta(H_0)},
\end{equation}
where we have included a normalization term in the denominator, $\beta(H_0)$, to ensure that the likelihood integrates to unity.
We can factor the numerator in Eq.~\ref{eq:likelihood} as:
\begin{align}
\label{eq:nlikelihood}
\int & p(d_{\rm GW}, d_{\rm EM}, D_L, \alpha, \delta, z | H_0)\,dD_L d\alpha d\delta dz  \nonumber\\ 
&= \int p(d_{\rm GW} | D_L, \alpha, \delta) p(d_{\rm EM}| z, \alpha, \delta) p(D_L | z, H_0) p_0(z, \alpha, \delta \mid H_0) dD_L d\alpha d\delta dz \nonumber\\
&= \int p(d_{\rm GW} | D_L, \alpha, \delta) p(d_{\rm EM}| z, \alpha, \delta) \delta(D_L - \hat{D_L}(z, H_0)) p_0(z, \alpha, \delta \mid H_0) dD_L d\alpha d\delta dz\nonumber\\
& = \int p(d_{\rm GW} | \hat{D_L}(z,H_0), \alpha, \delta) p(d_{\rm EM}| z, \alpha, \delta) p_0(z, \alpha, \delta \mid H_0) d\alpha d\delta dz,
\end{align}
where $\hat{D_L}(z,H_0)$ denotes the luminosity distance of a source at redshift $z$, given a Hubble constant of $H_0$ and leaving all other cosmological parameters fixed to the Planck values $(\Omega_{M_0}=0.308,\Omega_{\Lambda_0}=0.692)$ \cite{PlanckCosmology}. We can alternatively marginalize over these other cosmological parameters, but since most detected binaries will be at low redshifts, the effects of other cosmological parameters on the $z$--$D_L$ relation are small.
The term $p(d_{\rm GW} | D_L, \alpha, \delta)$ is the marginalized likelihood of the gravitational wave data given a compact binary source at distance $D_L$ and sky position ($\alpha$, $\delta$), marginalized over all other parameters.  
Throughout, we assume that we can construct a catalog of the potential host galaxies for each event, and take the prior $p_0(z, \alpha, \delta \mid H_0)$ to be
a sum of Gaussian distributions centered at the measured redshifts and sky positions of the galaxies:
\begin{align}
\label{eq:zprior}
p_0(z, \alpha, \delta \mid H_0) &= p_\mathrm{catalog}(z,\alpha,\delta) \nonumber\\ &= \frac{1}{N_{\rm gal}} \sum_i^{N_\mathrm{gal}} N[\bar{z}^i,\sigma_z^i](z)N[\bar{\alpha}^i,\sigma_\alpha^i](\alpha)N[\bar{\delta}^i,\sigma_\delta^i](\delta).
\end{align}
In practice, we ignore the uncertainties on the measured sky coordinates and treat the Gaussian distributions as $\delta$-functions centered at the measured $\bar{\alpha}^i$, $\bar{\delta}^i$. We take $\bar{z}^i$ to be the peculiar velocity-corrected redshifts, and assume a standard deviation of $c\sigma_z^i = 200$ km/s for each. 
In the above, we assign equal weights to each galaxy in the catalog, but if we believe that certain galaxies are a priori 
more likely to be GW hosts, we can assign weights accordingly. For example, we can weigh the galaxies in Eq.~\ref{eq:zprior} 
by their stellar or star-forming luminosity, or by some assumed redshift-dependent coalescence rate of the GW sources. 
A critical assumption is that the sum in Eq.~\ref{eq:zprior} contains the correct host galaxy. If we believe the catalog is 
incomplete, we must replace our prior, $p_0(z)$, with a weighted sum containing both the observed galaxies, 
Eq.~\ref{eq:zprior}---weighted by the overall completeness fraction of the catalog---and a smooth, 
uniform-in-comoving volume distribution accounting for the unobserved galaxies: 
\begin{align}
\label{eq:incompleteness}
p_0(z, \alpha, \delta \mid H_0) = f p_{\rm catalog}(z, \alpha, \delta) + (1-f)p_{\rm miss}(z, \alpha, \delta \mid H_0),
\end{align}
where $p_\mathrm{catalog}$ is given by Eq.~\ref{eq:zprior}, and:
\begin{equation}
p_\mathrm{miss}(z,\alpha,\delta \mid H_0) \propto \left[ 1-p_{\rm complete}(z, \alpha, \delta) \right] \frac{dV_c}{dz d\alpha d\delta},
\end{equation}
where $p_{\rm complete}(z, \alpha, \delta)$ is the probability of a galaxy at $(z, \alpha, \delta)$ being in the catalog.
Meanwhile the completeness fraction $f$ is given by:
\begin{equation}
f = \frac{1}{V_c(z_\mathrm{max})} {\int_\alpha \int_\delta \int_0^{z_\mathrm{max}} p_{\rm complete}(z, \alpha, \delta) \frac{dV_c}{dz d\alpha d\delta} dz d\alpha d \delta},
\end{equation}
where $z_{\rm max}$ is the maximum galaxy redshift considered in the analysis of an individual event, and $V_c(z_\mathrm{max})$ is the total comoving volume enclosed within $z_{\rm max}$.

In the case where we have an electromagnetic counterpart, the likelihood $p(d_{\rm EM} | z, \alpha, \delta)$ picks out one of the galaxies in the catalog, so that the sum in the prior reduces to a single term corresponding to the EM-identified host galaxy.
In the case where there is no electromagnetic counterpart, the EM data is uninformative, and we set the likelihood $p(d_{\rm EM} | z, \alpha, \delta) \propto$ constant.
In the case where we have an electromagnetic counterpart but cannot pick out a unique host galaxy, one could consider 
a ``pencil beam'' containing all the potential host galaxies within $\sim 100$ kpc in projected distance on the sky. 
We assume the sky position of the counterpart is perfectly measured to be $(\bar{\alpha},\bar{\delta})$, and take the term $p(d_{\rm EM} \mid z, \alpha, \delta)$ to be a top hat which picks out all of the galaxies within some angular radius, $r$, (corresponding to $\sim100$~kpc in projected distance) of the counterpart's sky position. Thus, the numerator of Eq.~\ref{eq:nlikelihood} reduces to:
\begin{equation}
\int_{\langle (\alpha,\delta) \mid (\bar{\alpha},\bar{\delta}) \rangle < r} p(d_{\rm GW} | \hat{D_L}(z,H_0), \bar{\alpha}, \bar{\delta}) p_0(z, \alpha, \delta) dz d\alpha d\delta,
\end{equation}
and we sum over all galaxies within $\sim100$~kpc in projected distance, but no longer weigh them by the likelihood of the GW source at the corresponding sky position. Alternatively, we can incorporate assumptions about the kick distribution in the form of $p(d_{EM} \mid z, \alpha, \delta)$ and place more weight at galaxies close to $(\bar{\alpha}, \bar{\delta})$ rather than assuming a simple top hat. Although for simplicity we don't apply the pencil-beam approach in this letter, it can be thought of as a natural interpolation between the counterpart and statistical cases.

To calculate the normalization term, $\beta(H_0)$, in the denominator of Eq.~\ref{eq:likelihood}, we must account for selection effects in our measurement process. In general, the GW and EM data are both subject to selection effects in that we only detect GW and EM sources that are above some threshold, $d_{\rm GW}^{\rm th}$ and $d_{\rm EM}^{\rm th}$, respectively. Accounting for these detection thresholds, the denominator of Eq.~\ref{eq:likelihood} is:
\begin{align}
\label{eq:beta}
\beta(H_0) = \int_{d_{\rm EM}>d_{\rm EM}^{\rm th}} \int_{d_{\rm GW}>d_{\rm GW}^{\rm th}} \int^4 p(d_{\rm GW}, d_{\rm EM}, D_L, \alpha, \delta, z | H_0) dD_L dz d\alpha d\delta d d_{\rm GW} d d_{\rm EM}
\end{align}
We define: 
\begin{equation}
p^{\rm GW}_{\rm det}(D_L,\alpha,\delta) \equiv \int_{d_{\rm GW}>d_{\rm GW}^{\rm th}}p(d_{\rm GW}|D_L,\alpha,\delta)dd_{\rm GW},
\end{equation}
and similarly:
\begin{equation}
p^{\rm EM}_{\rm det}(z, \alpha, \delta) \equiv \int_{d_{\rm EM}>d_{\rm EM}^{\rm th}}p(d_{\rm EM}|z, \alpha, \delta) dd_{\rm EM}.
\end{equation} 
With these definitions, Eq.~\ref{eq:beta} becomes:
\begin{align}
\label{eq:beta1}
&\nonumber\beta(H_0) = \int p^{\rm GW}_{\rm det}(\hat{D_L}(z,H_0),\alpha,\delta)p^{\rm EM}_{\rm det}(z,\alpha,\delta)p_0(z,\alpha,\delta) d\alpha d\delta dz.
\end{align}
Note that we have applied the same chain-rule factorization to the inner four integrals as in Eq.~\ref{eq:nlikelihood}.
It is clear that the  normalization factor $\beta(H_0)$ depends on $H_0$, so it is crucial to include it in the likelihood.
For the EM selection effects, we assume that we can detect all EM counterparts and host galaxies up to some maximum true 
redshift, $z_\mathrm{max}$. This is an over-simplification of the true EM selection effects, but is a reasonable assumption 
for the real-time galaxy catalogs that will be constructed during the EM follow-up to GW events. 
For example, at Advanced LIGO design sensitivity, 90\% of 30--30 \msun\ BBH detections will be within 5 Gpc 
(and BNS detections will be within 0.3 Gpc) \cite{Chen:distances}\footnote{http://gwc.rcc.uchicago.edu/}. 
Furthermore, only the BBHs with the smallest localization volumes contribute to the $H_0$ constraints, and these will typically be within 400 Mpc.
A galaxy with the same absolute magnitude as the Milky Way would have an apparent 
magnitude of $<23$ at typical 30--30 \msun\ BBH distances, or $<17.5$ for well-localized BBHs, 
and $<17$ at typical BNS distances. Meanwhile, we expect kilonova counterparts to BNS mergers to have magnitudes 
of $\leq 21.7$ on the first night, even at the farthest distances detectable by the HLVJI network at design sensitivity (assuming the optical counterpart to GW170817 is 
typical \cite{2017ApJ...848L..12A, 2017ApJ...848L..16S}).
This is well within the magnitude limits of upcoming survey telescopes (e.g. LSST), as well as within reach of current instruments (e.g. DECam, Subaru Hyper Suprime-Cam, ZTF etc.).  

We assume that EM counterparts are detectable for binaries regardless of the binary inclination. Although the short gamma-ray burst emission is expected to be beamed, the associated kilonovae are expected to emit isotropically. Furthermore, as GW170817 demonstrated, it is possible to identify a kilonova counterpart independently of the gamma-ray burst. We note that since face-on/face-off binaries are louder (have higher SNR) than edge-on binaries in GWs, the population of detected binaries will show a preference for face-on/face-off; our analysis reproduces the expected inclination distribution among detected sources (see Fig.~4 of \cite{2011CQGra..28l5023S}).
Under these assumptions for the EM selection effects, we have:
\begin{equation}
\label{eq:EMsel}
p^{\rm EM}_{\rm det}(z,\alpha,\delta) \propto \mathcal{H}(z_\mathrm{max}-z),
\end{equation}
where $\mathcal{H}$ is the Heaviside step function, and Eq.~\ref{eq:beta} reduces to:
\begin{equation}
\nonumber\beta(H_0) = 
\int_0^{z_\mathrm{max}}\int \int p^{\rm GW}_{\rm det}(\hat{D_L}(z,H_0),\alpha,\delta)p_0(z,\alpha,\delta) d\alpha d\delta dz.
\end{equation}

Meanwhile, we assume that the galaxy distribution is approximately isotropic on large scales, 
so that the galaxy catalog prior can be factored as:
\begin{equation}
p_0(z,\alpha,\delta) \approx p_0(z)p_0(\alpha,\delta),
\end{equation}
and we approximate $p_0(\alpha, \delta)$ by a continuous, isotropic distribution on the sky. We note that this assumption would only introduce systematic errors if the galaxy distribution had significant correlations with the sky sensitivities of the detectors, which is not to be expected.
We then define:
\begin{equation}
\label{eq:GWsel}
p^{\rm GW}_{\rm det}(D_L) = \int p^{\rm GW}_{\rm det}(D_L,\alpha,\delta)p_0(\alpha,\delta) d\alpha d\delta.
\end{equation}
We assume a detection threshold for GW sources corresponding to a matched-filter network SNR $\rho_{\rm th} = 12$, so that the detection probability, $p^{\rm GW}_{\rm det}(D_L)$, is the probability that a source at distance $D_L$ will have SNR $\rho > 12$. Assuming a distribution of orientations, masses and spins among a population of sources, in addition to the assumed isotropic distribution on the sky, $p_0(\alpha, \delta)$, we calculate the fraction of sources that are detectable at a given distance, $p^{\rm GW}_{\rm det}(D_L)$. We assume an isotropic distribution of orientation angles, and fix spins to be zero. For simplicity, we assume a monochromatic mass distribution for each type of source. (For example, we take all BNS sources to be 1.4--1.4 \msun.) 
We therefore have:
\begin{equation}
\label{eq:beta_final}
\beta(H_0) = \int_0^{z_\mathrm{max}} p^{\rm GW}_{\rm det}\left( D_L(z,H_0) \right) p_0(z)dz.
\end{equation}

We note that for GW sources in the local universe ($D_L<50$ Mpc), 
\begin{equation}
\label{eq:dl_local}
\hat{D_L}(z,H_0) \approx \frac{cz}{H_0}.
\end{equation}
If we assume that the distribution of galaxies is uniform in comoving volume, then in the local universe, we can approximate:
\begin{equation}
\label{eq:zprior_local}
p_0(z) \propto z^2.
\end{equation}
With these approximations, assuming that EM selection effects are negligible ($z_\mathrm{max} \rightarrow \infty$), $\beta(H_0)$ is independent of the masses of the source, which determine the distance to which it can be detected. In fact, under these assumptions, $\beta(H_0)$ simplifies to:
\begin{equation}
\beta(H_0) \propto H_0^3.
\end{equation}
However, in general, we must account for cosmological deviations from Eqs.~\ref{eq:dl_local} and \ref{eq:zprior_local} so we calculate $\beta(H_0)$ according to Eq.~\ref{eq:beta_final} throughout our analysis. We note that $\beta(H_0)$ is still only weakly dependent on the GW horizon and therefore the (unknown) underlying mass distribution of GW sources. Nevertheless, the statistical framework described here can accommodate more complicated models of the GW source distribution and its effects on the detection probability (Eq.~\ref{eq:GWsel}). 

\begin{figure}
\centering
\includegraphics[width=1.1\columnwidth]{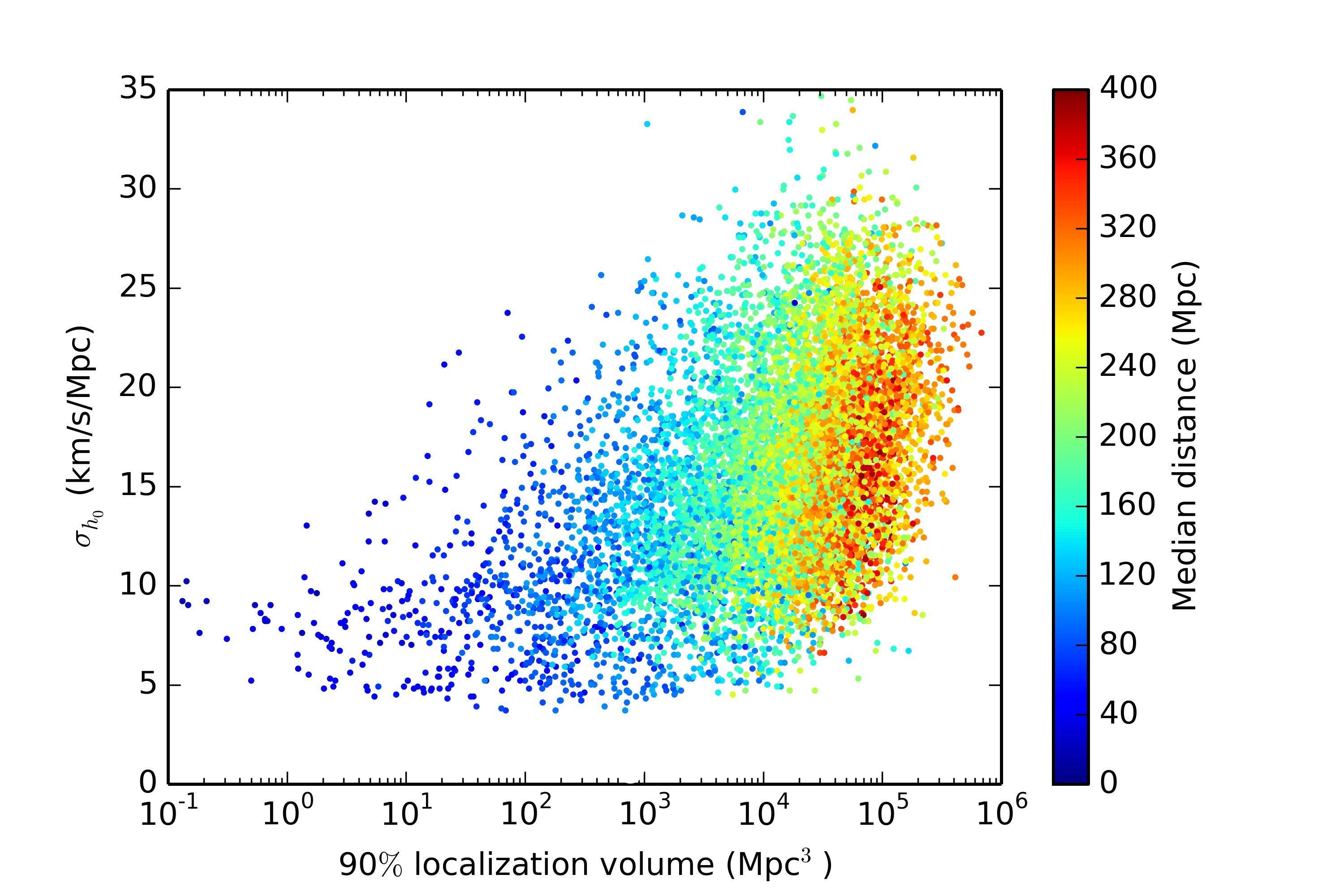}
\captionsetup{labelformat=empty}
\caption{
\textbf{(Extended Data) $H_0$ uncertainty $\sigma_{h_0}$ for BNS systems with identified counterparts and redshifts} Each point is the $H_0$ uncertainty from a simulated detection with the Advanced LIGO-Hanford+LIGO-Livingston+Virgo network operating at design sensitivity, as a function of the 90$\%$ localization volume. The colors correspond to the median of the gravitational-wave distance measurement. The lower $\sim3$ km/s/Mpc limit to the precision of individual measurements is due to the ``sweet spot'' between peculiar velocities and distance uncertainties, as discussed in the text. We find that, in general, closer events have smaller localization volume and lead to better constraints on $H_0$, although the closest events yield slightly worse constraints because of their larger fractional peculiar velocity uncertainties.
}
\end{figure}

\bibliography{references}

\begin{thebibliography}{10}
\expandafter\ifx\csname url\endcsname\relax
  \def\url#1{\texttt{#1}}\fi
\expandafter\ifx\csname urlprefix\endcsname\relax\def\urlprefix{URL }\fi
\providecommand{\bibinfo}[2]{#2}
\providecommand{\eprint}[2][]{\url{#2}}

\bibitem{Schutz1986}
\bibinfo{author}{{Schutz}, B.~F.}
\newblock \bibinfo{title}{{Determining the Hubble constant from gravitational
  wave observations}}.
\newblock \emph{\bibinfo{journal}{\nat}} \textbf{\bibinfo{volume}{323}},
  \bibinfo{pages}{310--+} (\bibinfo{year}{1986}).

\bibitem{Holz2005}
\bibinfo{author}{Holz, D.~E.} \& \bibinfo{author}{Hughes, S.~A.}
\newblock \bibinfo{title}{Using gravitational-wave standard sirens}.
\newblock \emph{\bibinfo{journal}{\apj}} \textbf{\bibinfo{volume}{629}},
  \bibinfo{pages}{15} (\bibinfo{year}{2005}).
\newblock \urlprefix\url{http://stacks.iop.org/0004-637X/629/i=1/a=15}.

\bibitem{2006PhRvD..74f3006D}
\bibinfo{author}{{Dalal}, N.}, \bibinfo{author}{{Holz}, D.~E.},
  \bibinfo{author}{{Hughes}, S.~A.} \& \bibinfo{author}{{Jain}, B.}
\newblock \bibinfo{title}{{Short GRB and binary black hole standard sirens as a
  probe of dark energy}}.
\newblock \emph{\bibinfo{journal}{\prd}} \textbf{\bibinfo{volume}{74}},
  \bibinfo{pages}{063006} (\bibinfo{year}{2006}).
\newblock \eprint{astro-ph/0601275}.

\bibitem{2017Natur.551...85A}
\bibinfo{author}{{Abbott}, B.~P.} \emph{et~al.}
\newblock \bibinfo{title}{{A gravitational-wave standard siren measurement of
  the Hubble constant}}.
\newblock \emph{\bibinfo{journal}{\nat}} \textbf{\bibinfo{volume}{551}},
  \bibinfo{pages}{85--88} (\bibinfo{year}{2017}).
\newblock \eprint{1710.05835}.

\bibitem{SHoES}
\bibinfo{author}{{Riess}, A.~G.} \emph{et~al.}
\newblock \bibinfo{title}{{A 2.4\% Determination of the Local Value of the
  Hubble Constant}}.
\newblock \emph{\bibinfo{journal}{\apj}} \textbf{\bibinfo{volume}{826}},
  \bibinfo{pages}{56} (\bibinfo{year}{2016}).
\newblock \eprint{1604.01424}.

\bibitem{PlanckCosmology}
\bibinfo{author}{{Ade}, P.~A.~R.} \emph{et~al.}
\newblock \bibinfo{title}{{Planck 2015 results. XIII. Cosmological
  parameters}}.
\newblock \emph{\bibinfo{journal}{\aap}} \textbf{\bibinfo{volume}{594}},
  \bibinfo{pages}{A13} (\bibinfo{year}{2016}).
\newblock \eprint{1502.01589}.

\bibitem{Riess:2018}
\bibinfo{author}{{Riess}, A.~G.} \emph{et~al.}
\newblock \bibinfo{title}{{New Parallaxes of Galactic Cepheids from Spatially
  Scanning the Hubble Space Telescope: Implications for the Hubble Constant}}.
\newblock \emph{\bibinfo{journal}{ArXiv e-prints}}  (\bibinfo{year}{2018}).
\newblock \eprint{1801.01120}.

\bibitem{2005ASPC..339..215H}
\bibinfo{author}{{Hu}, W.}
\newblock \bibinfo{title}{{Dark Energy Probes in Light of the CMB}}.
\newblock In \bibinfo{editor}{{Wolff}, S.~C.} \& \bibinfo{editor}{{Lauer},
  T.~R.} (eds.) \emph{\bibinfo{booktitle}{Observing Dark Energy}}, vol.
  \bibinfo{volume}{339} of \emph{\bibinfo{series}{Astronomical Society of the
  Pacific Conference Series}}, \bibinfo{pages}{215} (\bibinfo{year}{2005}).
\newblock \eprint{astro-ph/0407158}.

\bibitem{2018arXiv180607463D}
\bibinfo{author}{{Di Valentino}, E.}, \bibinfo{author}{{Holz}, D.~E.},
  \bibinfo{author}{{Melchiorri}, A.} \& \bibinfo{author}{{Renzi}, F.}
\newblock \bibinfo{title}{{The cosmological impact of future constraints on
  $H\_0$ from gravitational-wave standard sirens}}.
\newblock \emph{\bibinfo{journal}{ArXiv e-prints}}  (\bibinfo{year}{2018}).
\newblock \eprint{1806.07463}.

\bibitem{1998ApJ...507L..59L}
\bibinfo{author}{{Li}, L.-X.} \& \bibinfo{author}{{Paczy{\'n}ski}, B.}
\newblock \bibinfo{title}{{Transient Events from Neutron Star Mergers}}.
\newblock \emph{\bibinfo{journal}{\apjl}} \textbf{\bibinfo{volume}{507}},
  \bibinfo{pages}{L59--L62} (\bibinfo{year}{1998}).
\newblock \eprint{astro-ph/9807272}.

\bibitem{2010MNRAS.406.2650M}
\bibinfo{author}{{Metzger}, B.~D.} \emph{et~al.}
\newblock \bibinfo{title}{{Electromagnetic counterparts of compact object
  mergers powered by the radioactive decay of r-process nuclei}}.
\newblock \emph{\bibinfo{journal}{\mnras}} \textbf{\bibinfo{volume}{406}},
  \bibinfo{pages}{2650--2662} (\bibinfo{year}{2010}).
\newblock \eprint{1001.5029}.

\bibitem{2017Sci...358.1556C}
\bibinfo{author}{{Coulter}, D.~A.} \emph{et~al.}
\newblock \bibinfo{title}{{Swope Supernova Survey 2017a (SSS17a), the optical
  counterpart to a gravitational wave source}}.
\newblock \emph{\bibinfo{journal}{Science}} \textbf{\bibinfo{volume}{358}},
  \bibinfo{pages}{1556--1558} (\bibinfo{year}{2017}).
\newblock \eprint{1710.05452}.

\bibitem{2017ApJ...848L..16S}
\bibinfo{author}{{Soares-Santos}, M.} \emph{et~al.}
\newblock \bibinfo{title}{{The Electromagnetic Counterpart of the Binary
  Neutron Star Merger LIGO/Virgo GW170817. I. Discovery of the Optical
  Counterpart Using the Dark Energy Camera}}.
\newblock \emph{\bibinfo{journal}{\apjl}} \textbf{\bibinfo{volume}{848}},
  \bibinfo{pages}{L16} (\bibinfo{year}{2017}).
\newblock \eprint{1710.05459}.

\bibitem{2013ApJ...776...18F}
\bibinfo{author}{{Fong}, W.} \& \bibinfo{author}{{Berger}, E.}
\newblock \bibinfo{title}{{The Locations of Short Gamma-Ray Bursts as Evidence
  for Compact Object Binary Progenitors}}.
\newblock \emph{\bibinfo{journal}{\apj}} \textbf{\bibinfo{volume}{776}},
  \bibinfo{pages}{18} (\bibinfo{year}{2013}).
\newblock \eprint{1307.0819}.

\bibitem{2010PhRvD..81l4046H}
\bibinfo{author}{{Hirata}, C.~M.}, \bibinfo{author}{{Holz}, D.~E.} \&
  \bibinfo{author}{{Cutler}, C.}
\newblock \bibinfo{title}{{Reducing the weak lensing noise for the
  gravitational wave Hubble diagram using the non-Gaussianity of the
  magnification distribution}}.
\newblock \emph{\bibinfo{journal}{\prd}} \textbf{\bibinfo{volume}{81}}
  (\bibinfo{year}{2010}).

\bibitem{2013arXiv1307.2638N}
\bibinfo{author}{{Nissanke}, S.} \emph{et~al.}
\newblock \bibinfo{title}{{Determining the Hubble constant from gravitational
  wave observations of merging compact binaries}}.
\newblock \emph{\bibinfo{journal}{ArXiv e-prints}}  (\bibinfo{year}{2013}).
\newblock \eprint{1307.2638}.

\bibitem{Abbott:170817}
\bibinfo{author}{Abbott, B.~P.} \emph{et~al.}
\newblock \bibinfo{title}{Gw170817: Observation of gravitational waves from a
  binary neutron star inspiral}.
\newblock \emph{\bibinfo{journal}{Phys. Rev. Lett.}}
  \textbf{\bibinfo{volume}{119}}, \bibinfo{pages}{161101}
  (\bibinfo{year}{2017}).
\newblock
  \urlprefix\url{https://link.aps.org/doi/10.1103/PhysRevLett.119.161101}.

\bibitem{Abbott:170104}
\bibinfo{author}{{Abbott}, B.~P.} \emph{et~al.}
\newblock \bibinfo{title}{Gw170104: Observation of a 50-solar-mass binary black
  hole coalescence at redshift 0.2}.
\newblock \emph{\bibinfo{journal}{Phys. Rev. Lett.}}
  \textbf{\bibinfo{volume}{118}}, \bibinfo{pages}{221101}
  (\bibinfo{year}{2017}).
\newblock
  \urlprefix\url{https://link.aps.org/doi/10.1103/PhysRevLett.118.221101}.

\bibitem{MDC}
\bibinfo{author}{{Gray}, R.} \emph{et~al.}
\newblock \bibinfo{title}{Simulating the gravitational-wave cosmology}
  (\bibinfo{year}{2018}).

\bibitem{Dominik:2015}
\bibinfo{author}{{Dominik}, M.} \emph{et~al.}
\newblock \bibinfo{title}{{Double Compact Objects III: Gravitational-wave
  Detection Rates}}.
\newblock \emph{\bibinfo{journal}{\apj}} \textbf{\bibinfo{volume}{806}},
  \bibinfo{pages}{263} (\bibinfo{year}{2015}).
\newblock \eprint{1405.7016}.

\bibitem{Belczynski:2016b}
\bibinfo{author}{{Belczynski}, K.} \emph{et~al.}
\newblock \bibinfo{title}{{Compact Binary Merger Rates: Comparison with
  LIGO/Virgo Upper Limits}}.
\newblock \emph{\bibinfo{journal}{\apj}} \textbf{\bibinfo{volume}{819}},
  \bibinfo{pages}{108} (\bibinfo{year}{2016}).
\newblock \eprint{1510.04615}.

\bibitem{2016Natur.534..512B}
\bibinfo{author}{{Belczynski}, K.}, \bibinfo{author}{{Holz}, D.~E.},
  \bibinfo{author}{{Bulik}, T.} \& \bibinfo{author}{{O'Shaughnessy}, R.}
\newblock \bibinfo{title}{{The first gravitational-wave source from the
  isolated evolution of two stars in the 40-100 solar mass range}}.
\newblock \emph{\bibinfo{journal}{\nat}} \textbf{\bibinfo{volume}{534}},
  \bibinfo{pages}{512--515} (\bibinfo{year}{2016}).
\newblock \eprint{1602.04531}.

\bibitem{2018arXiv180407337V}
\bibinfo{author}{{Vitale}, S.} \& \bibinfo{author}{{Chen}, H.-Y.}
\newblock \bibinfo{title}{{Measuring the Hubble constant with neutron star
  black hole mergers}}.
\newblock \emph{\bibinfo{journal}{ArXiv e-prints}}  (\bibinfo{year}{2018}).
\newblock \eprint{1804.07337}.

\bibitem{2017arXiv171006426G}
\bibinfo{author}{{Guidorzi}, C.} \emph{et~al.}
\newblock \bibinfo{title}{{Improved constraints on H0 from a combined analysis
  of gravitational-wave and electromagnetic emission from GW170817}}.
\newblock \emph{\bibinfo{journal}{ArXiv e-prints}}  (\bibinfo{year}{2017}).
\newblock \eprint{1710.06426}.

\bibitem{calibration}
\bibinfo{author}{{Karki}, S.} \emph{et~al.}
\newblock \bibinfo{title}{{The Advanced LIGO photon calibrators}}.
\newblock \emph{\bibinfo{journal}{Review of Scientific Instruments}}
  \textbf{\bibinfo{volume}{87}}, \bibinfo{pages}{114503}
  (\bibinfo{year}{2016}).
\newblock \eprint{1608.05055}.

\bibitem{2005ApJ...631..678H}
\bibinfo{author}{{Holz}, D.~E.} \& \bibinfo{author}{{Linder}, E.~V.}
\newblock \bibinfo{title}{{Safety in Numbers: Gravitational Lensing Degradation
  of the Luminosity Distance-Redshift Relation}}.
\newblock \emph{\bibinfo{journal}{\apj}} \textbf{\bibinfo{volume}{631}},
  \bibinfo{pages}{678--688} (\bibinfo{year}{2005}).
\newblock \eprint{astro-ph/0412173}.

\bibitem{2016ApJ...833L...1A}
\bibinfo{author}{{Abbott}, B.~P.} \emph{et~al.}
\newblock \bibinfo{title}{{The Rate of Binary Black Hole Mergers Inferred from
  Advanced LIGO Observations Surrounding GW150914}}.
\newblock \emph{\bibinfo{journal}{\apjl}} \textbf{\bibinfo{volume}{833}},
  \bibinfo{pages}{L1} (\bibinfo{year}{2016}).
\newblock \eprint{1602.03842}.

\bibitem{2017arXiv170908584F}
\bibinfo{author}{{Fishbach}, M.} \& \bibinfo{author}{{Holz}, D.~E.}
\newblock \bibinfo{title}{{Where are LIGO's Big Black Holes?}}
\newblock \emph{\bibinfo{journal}{\apjl}} \textbf{\bibinfo{volume}{851}},
  \bibinfo{pages}{L25} (\bibinfo{year}{2017}).
\newblock \eprint{1709.08584}.

\bibitem{Chen:theone}
\bibinfo{author}{{Chen}, H.-Y.} \& \bibinfo{author}{{Holz}, D.~E.}
\newblock \bibinfo{title}{{Finding the One: Identifying the Host Galaxies of
  Gravitational-Wave Sources}}.
\newblock \emph{\bibinfo{journal}{ArXiv e-prints}}  (\bibinfo{year}{2016}).
\newblock \eprint{1612.01471}.

\bibitem{Abbott:obs}
\bibinfo{author}{{Abbott}, B.~P.} \emph{et~al.}
\newblock \bibinfo{title}{{Prospects for Observing and Localizing
  Gravitational-Wave Transients with Advanced LIGO and Advanced Virgo}}.
\newblock \emph{\bibinfo{journal}{Living Reviews in Relativity}}
  \textbf{\bibinfo{volume}{19}}, \bibinfo{pages}{1} (\bibinfo{year}{2016}).
\newblock \eprint{1304.0670}.

\bibitem{Veitch:2015}
\bibinfo{author}{{Veitch}, J.} \emph{et~al.}
\newblock \bibinfo{title}{{Parameter estimation for compact binaries with
  ground-based gravitational-wave observations using the LALInference software
  library}}.
\newblock \emph{\bibinfo{journal}{\prd}} \textbf{\bibinfo{volume}{91}},
  \bibinfo{pages}{042003} (\bibinfo{year}{2015}).
\newblock \eprint{1409.7215}.

\bibitem{Carrick:2015}
\bibinfo{author}{{Carrick}, J.}, \bibinfo{author}{{Turnbull}, S.~J.},
  \bibinfo{author}{{Lavaux}, G.} \& \bibinfo{author}{{Hudson}, M.~J.}
\newblock \bibinfo{title}{{Cosmological parameters from the comparison of
  peculiar velocities with predictions from the 2M++ density field}}.
\newblock \emph{\bibinfo{journal}{\mnras}} \textbf{\bibinfo{volume}{450}},
  \bibinfo{pages}{317--332} (\bibinfo{year}{2015}).
\newblock \eprint{1504.04627}.

\bibitem{Scolnic:2017b}
\bibinfo{author}{{Scolnic}, D.~M.} \emph{et~al.}
\newblock \bibinfo{title}{{The Complete Light-curve Sample of Spectroscopically
  Confirmed Type Ia Supernovae from Pan-STARRS1 and Cosmological Constraints
  from The Combined Pantheon Sample}}.
\newblock \emph{\bibinfo{journal}{ArXiv e-prints}}  (\bibinfo{year}{2017}).
\newblock \eprint{1710.00845}.

\bibitem{2012PhRvD..86d3011D}
\bibinfo{author}{{Del Pozzo}, W.}
\newblock \bibinfo{title}{{Inference of cosmological parameters from
  gravitational waves: Applications to second generation interferometers}}.
\newblock \emph{\bibinfo{journal}{\prd}} \textbf{\bibinfo{volume}{86}},
  \bibinfo{pages}{043011} (\bibinfo{year}{2012}).
\newblock \eprint{1108.1317}.

\bibitem{2015MNRAS.448.2987F}
\bibinfo{author}{{Fosalba}, P.}, \bibinfo{author}{{Crocce}, M.},
  \bibinfo{author}{{Gazta{\~n}aga}, E.} \& \bibinfo{author}{{Castander}, F.~J.}
\newblock \bibinfo{title}{{The MICE grand challenge lightcone simulation - I.
  Dark matter clustering}}.
\newblock \emph{\bibinfo{journal}{\mnras}} \textbf{\bibinfo{volume}{448}},
  \bibinfo{pages}{2987--3000} (\bibinfo{year}{2015}).
\newblock \eprint{1312.1707}.

\bibitem{2015MNRAS.453.1513C}
\bibinfo{author}{{Crocce}, M.}, \bibinfo{author}{{Castander}, F.~J.},
  \bibinfo{author}{{Gazta{\~n}aga}, E.}, \bibinfo{author}{{Fosalba}, P.} \&
  \bibinfo{author}{{Carretero}, J.}
\newblock \bibinfo{title}{{The MICE Grand Challenge lightcone simulation - II.
  Halo and galaxy catalogues}}.
\newblock \emph{\bibinfo{journal}{\mnras}} \textbf{\bibinfo{volume}{453}},
  \bibinfo{pages}{1513--1530} (\bibinfo{year}{2015}).
\newblock \eprint{1312.2013}.

\bibitem{2015MNRAS.447.1319F}
\bibinfo{author}{{Fosalba}, P.}, \bibinfo{author}{{Gazta{\~n}aga}, E.},
  \bibinfo{author}{{Castander}, F.~J.} \& \bibinfo{author}{{Crocce}, M.}
\newblock \bibinfo{title}{{The MICE Grand Challenge light-cone simulation -
  III. Galaxy lensing mocks from all-sky lensing maps}}.
\newblock \emph{\bibinfo{journal}{\mnras}} \textbf{\bibinfo{volume}{447}},
  \bibinfo{pages}{1319--1332} (\bibinfo{year}{2015}).
\newblock \eprint{1312.2947}.

\bibitem{Carretero:2017zkw}
\bibinfo{author}{Carretero, J.} \emph{et~al.}
\newblock \bibinfo{title}{{CosmoHub and SciPIC: Massive cosmological data
  analysis, distribution and generation using a Big Data platform}}.
\newblock \emph{\bibinfo{journal}{PoS}} \textbf{\bibinfo{volume}{EPS-HEP2017}},
  \bibinfo{pages}{488} (\bibinfo{year}{2017}).

\bibitem{1976ApJ...203..297S}
\bibinfo{author}{{Schechter}, P.}
\newblock \bibinfo{title}{{An analytic expression for the luminosity function
  for galaxies.}}
\newblock \emph{\bibinfo{journal}{\apj}} \textbf{\bibinfo{volume}{203}},
  \bibinfo{pages}{297--306} (\bibinfo{year}{1976}).

\bibitem{2002MNRAS.336..907N}
\bibinfo{author}{{Norberg}, P.} \emph{et~al.}
\newblock \bibinfo{title}{{The 2dF Galaxy Redshift Survey: the b$_{J}$-band
  galaxy luminosity function and survey selection function}}.
\newblock \emph{\bibinfo{journal}{\mnras}} \textbf{\bibinfo{volume}{336}},
  \bibinfo{pages}{907--931} (\bibinfo{year}{2002}).
\newblock \eprint{astro-ph/0111011}.

\bibitem{2003MNRAS.344..307L}
\bibinfo{author}{{Liske}, J.}, \bibinfo{author}{{Lemon}, D.~J.},
  \bibinfo{author}{{Driver}, S.~P.}, \bibinfo{author}{{Cross}, N.~J.~G.} \&
  \bibinfo{author}{{Couch}, W.~J.}
\newblock \bibinfo{title}{{The Millennium Galaxy Catalogue: 16 <=B$_{MGC}$ < 24
  galaxy counts and the calibration of the local galaxy luminosity function}}.
\newblock \emph{\bibinfo{journal}{\mnras}} \textbf{\bibinfo{volume}{344}},
  \bibinfo{pages}{307--324} (\bibinfo{year}{2003}).
\newblock \eprint{astro-ph/0207555}.

\bibitem{2006A&A...445...51G}
\bibinfo{author}{{Gonz{\'a}lez}, R.~E.}, \bibinfo{author}{{Lares}, M.},
  \bibinfo{author}{{Lambas}, D.~G.} \& \bibinfo{author}{{Valotto}, C.}
\newblock \bibinfo{title}{{The faint-end of the galaxy luminosity function in
  groups}}.
\newblock \emph{\bibinfo{journal}{\aap}} \textbf{\bibinfo{volume}{445}},
  \bibinfo{pages}{51--58} (\bibinfo{year}{2006}).
\newblock \eprint{astro-ph/0507144}.

\bibitem{2017ApJ...843...16K}
\bibinfo{author}{{Kourkchi}, E.} \& \bibinfo{author}{{Tully}, R.~B.}
\newblock \bibinfo{title}{{Galaxy Groups Within 3500 km s$^{-1}$}}.
\newblock \emph{\bibinfo{journal}{\apj}} \textbf{\bibinfo{volume}{843}},
  \bibinfo{pages}{16} (\bibinfo{year}{2017}).
\newblock \eprint{1705.08068}.

\bibitem{Chen:distances}
\bibinfo{author}{{Chen}, H.-Y.} \emph{et~al.}
\newblock \bibinfo{title}{{Distance measures in gravitational-wave astrophysics
  and cosmology}}.
\newblock \emph{\bibinfo{journal}{ArXiv e-prints}}  (\bibinfo{year}{2017}).
\newblock \eprint{1709.08079}.

\bibitem{2017ApJ...848L..12A}
\bibinfo{author}{{Abbott}, B.~P.} \emph{et~al.}
\newblock \bibinfo{title}{{Multi-messenger Observations of a Binary Neutron
  Star Merger}}.
\newblock \emph{\bibinfo{journal}{\apjl}} \textbf{\bibinfo{volume}{848}},
  \bibinfo{pages}{L12} (\bibinfo{year}{2017}).
\newblock \eprint{1710.05833}.

\bibitem{2011CQGra..28l5023S}
\bibinfo{author}{{Schutz}, B.~F.}
\newblock \bibinfo{title}{{Networks of gravitational wave detectors and three
  figures of merit}}.
\newblock \emph{\bibinfo{journal}{Classical and Quantum Gravity}}
  \textbf{\bibinfo{volume}{28}}, \bibinfo{pages}{125023}
  (\bibinfo{year}{2011}).
\newblock \eprint{1102.5421}.

\end{thebibliography}

\end{document}